\documentclass[a4paper,fleqn,usenatbib]{mnras}

\usepackage{amsmath,latexsym}    % need for subequations
\usepackage[dvipdfmx]{graphicx}   % need for figures
\usepackage{verbatim}   % useful for program listings
\usepackage{color}      % use if color is used in text
\usepackage{subfigure}  % use for side-by-side figures
\usepackage{hyperref}   % use for hypertext links, including those to external documents and URLs
\usepackage{tensor}     %use for tensor, as it should be.
\usepackage{amssymb}
\usepackage{enumerate}	% can be used to enumerate several items
\usepackage{braket}		% for braket notation
\usepackage{newtxtext,newtxmath}

\newcommand{\mpcoh}{\,h^{-1}\,{\rm Mpc}}
\newcommand{\GR}{\rm GR}
\newcommand{\GG}{\rm GG}
\newcommand{\GI}{\rm GI}
\newcommand{\II}{\rm II}
\newcommand{\be}{\begin{equation}}
\newcommand{\ee}{\end{equation}}
\newcommand{\bs}{\rm bs}
\newcommand{\wt}{\widetilde}

\newcommand{\bfk}{\boldsymbol{k}}
\newcommand{\bfr}{\boldsymbol{r}}
\newcommand{\bfx}{\boldsymbol{x}}

\begin{document}

\title[
Halo ellipticity correlations in $f(R)$ simulations]
{
Distinguishing between $\Lambda$CDM and $f(R)$ gravity models using halo ellipticity correlations in simulations
}

\author[Y. T. Chuang, T. Okumura and M. Shirasaki]{
Yao-Tsung Chuang$^{1,2}$\thanks{ytchuang@asiaa.sinica.edu.tw},
Teppei Okumura$^{1,3}$\thanks{tokumura@asiaa.sinica.edu.tw} and 
Masato Shirasaki$^{4,5}$
\vspace*{2pt} %\\
\\
$^{1}$ Institute of Astronomy and Astrophysics, Academia Sinica, No. 1, Section 4, Roosevelt Road, Taipei 10617, Taiwan \\
$^{2}$ Department of Physics, National Taiwan University, No. 1, Section 4, Roosevelt Road, Taipei 10617, Taiwan \\
$^{3}$ Kavli Institute for the Physics and Mathematics of the Universe (WPI), UTIAS, The University of Tokyo, Kashiwa, Chiba 277-8583, Japan \\
$^{4}$ National Astronomical Observatory of Japan (NAOJ), Mitaka, Tokyo 181-8588, Japan \\
$^{5}$ The Institute of Statistical Mathematics, Tachikawa, Tokyo 190-8562, Japan
}
%\date{Accepted XXX. Received YYY; in original form ZZZ}
\date{\today} 
\pagerange{\pageref{firstpage}--\pageref{lastpage}} \pubyear{2021}

\maketitle
\label{firstpage}

\begin{abstract}
There is a growing interest in utilizing intrinsic alignment (IA) of
galaxy shapes as a geometric and dynamical probe of cosmology.  In
this paper we present the first measurements of IA in a modified
gravity model using the gravitational shear-intrinsic ellipticity
correlation (GI) and intrinsic ellipticity-ellipticity correlation
(II) functions of dark-matter halos from $f(R)$ gravity
simulations. By comparing them with the same statistics measured in
$\Lambda$CDM simulations, we find that the IA statistics in different
gravity models show distinguishable features, with a trend similar to
the case of conventional galaxy clustering statistics. Thus, the GI
and II correlations are found to be useful in distinguishing between
the $\Lambda$CDM and $f(R)$ gravity models. More quantitatively, IA
statistics enhance detectability of the imprint of $f(R)$ gravity on
large scale structures by $\sim 40\%$ when combined with the
conventional halo clustering in redshift space. 
We also find that the correlation between the axis ratio and orientation of halos becomes stronger in $f(R)$ gravity than that in $\Lambda$CDM. 
Our results
demonstrate the usefulness of IA statistics as a probe of gravity
beyond a consistency test of $\Lambda$CDM and general relativity.

\end{abstract}



\section{Introduction} \label{sec:intro}

The origin of the cosmic acceleration has been one of the most
profound mysteries for decades \citep{Weinberg:2013}. Many studies
have explored it by considering dark energy as a source of the cosmic
acceleration.  Modifying the law of gravity at cosmological scales is
an alternative way to explain the acceleration
\citep{Wang:1998,Linder:2005}.  Conventionally, galaxy clustering
observed in redshift surveys has been extensively exploited for this
purpose \citep[e.g.,][]{Guzzo:2008,Reyes:2010,Okumura:2016}.

Intrinsic alignment (IA) of galaxy shapes, originally focused as a
contaminant to gravitational lensing signals
\citep{Croft:2000gz,Heavens:2000,Hirata:2004,Mandelbaum:2006,Hirata:2007,Okumura:2009,Okumura:2009a,Blazek:2011,Troxel:2015,Tonegawa:2022},
has been drawing attention as a new dynamical and geometric probe of
cosmology
\cite[e.g.,][]{Chisari:2013,Okumura:2019ozd,Taruya:2020,Kurita:2021,Okumura:2021a,Reischke:2022}.
However, such a possibility has been explored merely by forecast
studies or numerical simulations based on the $\Lambda$CDM model.
Thus, we still do not know how the observed IA looks like in gravity
models beyond general relativity (GR).

In this paper, we present the first measurements of IA statistics of
dark-matter halos in modified gravity using N-body simulations of the
$f(R)$ gravity model. We then show that the IA statistics are indeed
useful to tighten the constraint on gravity models by combining with
the conventional galaxy clustering statistics.

This paper is organized as follows. In Section \ref{sec:IAtheory}, we briefly review the statistics of IA used in this paper. 
Section \ref{sec:f(R)gravity} describes N-body simulations under the $\Lambda$CDM and $f(R)$ gravity models. 
In Section \ref{sec:cf}, we present measurements of the halo clustering and alignment statistics. 
We investigate how well one can improve the distinguishability between the $\Lambda$CDM and $f(R)$ gravity models by considering the IA statistics in section \ref{sec:results}. 
Our conclusions are given in Section \ref{sec:conclusion}.

\section{Intrinsic alignment statistics} \label{sec:IAtheory}
This section provides a brief description of the three-dimensional
alignment statistics following \citet{Okumura:2020}.  In this paper,
we measure all the statistics in redshift space, and thus the halo
overdensity field $\delta_h$ below is sampled in redshift space and
suffers from redshift-space distortions \citep[RSD][]{Kaiser:1987}.

To begin with, orientations of halos or galaxies projected onto the
sky are quantified by the two-component ellipticity, given as
\begin{equation}
\label{eq:gamma}
\gamma_{(+,\times)}(\bfx)=\frac{1-q^2}{1+q^2}(\cos(2\theta),\sin(2\theta)),
\end{equation}
where $\theta$, defined on the plane normal to the line-of-sight, is
the angle between the major axis projected onto the celestial sphere
and the projected separation vector pointing to another object,
$q$ is the minor-to-major axis ratio on the projected plane ($0\leq q \leq 1$).

In this paper, together with the halo density correlation function,
$\xi_{hh}$, abbreviated as the GG correlation, we study two types of
IA statistics, the intrinsic ellipticity (II) correlation functions,
$\xi_{+}$ and $\xi_{-}$ \citep{Croft:2000gz,Heavens:2000}, and the
gravitational shear-intrinsic ellipticity (GI) correlation functions,
$\xi_{h+}$ \citep{Hirata:2004}. These IA statistics are concisely defined as
\begin{align}
    \xi_{X}(\bfr) = \langle[1+\delta_{h}(\bfx_{1})][1+\delta_{h}(\bfx_{2})]W_X(\bfx_{1},\bfx_{2})\rangle, \label{eq:statistics}
\end{align}
where $X = \{h+,+,-\}$ and ${\bf r}={\bf x}_2-{\bf x}_1$. The GI and II
correlation functions are characterized by the function
$W_X(\bfx_{1},\bfx_{2})$: 
$W_{h+}(\bfx_{1},\bfx_{2})=\gamma_+(\bfx_2)$ and
$W_{\pm}(\bfx_{1},\bfx_{2})=\gamma_+(\bfx_1)\gamma_+(\bfx_2)\pm
\gamma_\times(\bfx_1)\gamma_\times(\bfx_2)$ for the GI and II
correlation functions, respectively.  For the II correlation, when we
specifically refer to $\xi_{+}$ and $\xi_{-}$, they are abbreviated as
the II($+$) and II($-$) correlations, respectively.  Throughout this
paper, we assume the distant-observer approximation so that
$\hat{\bfx}_{1}=\hat{\bfx}_{2}\equiv \hat{\bfx}$, where a hat denotes
a unit vector.

\section{$f(R)$ gravity simulations} \label{sec:f(R)gravity}
In $f(R)$ gravity theories, we replace the Ricci scalar $R$ in the
Einstein-Hilbert action by some general function of $R$, $f(R)$, to
mimic the effect of the cosmological constant $\Lambda$
\citep{Starobinsky:1980,De_Felice:2010,Nojiri:2011}.  We adopt the functional form
of $f(R)$ introduced by \cite{Hu:2007nk}, where the deviation of the
law of gravity from GR is characterized by $f_{R0}\equiv
df(R)/dR|_{z=0}$.

We perform $N$-body simulations under the $\Lambda$CDM model and
$f(R)$ gravity and study the difference of the alignment statistics
measured from them.  The cosmological N-body simulations have been run
with ECOSMOG code \citep{2012JCAP...01..051L}.  Every simulation in
our paper consists of $512^3$ particles in a cubic box with the side
length being $L = 316 h^{-1}\mathrm{Mpc}$.  The initial condition of
the simulation is generated with the 2LPTic code using second-order
Lagrangian perturbation theory \citep{2006MNRAS.373..369C}.  We set
the initial redshift to $z=50$ and mainly work with the simulation
outputs at $z=0$.  Note that the same initial condition has been
used for both of $\Lambda$CDM model and $f(R)$ gravity runs.  Each
simulation assumes the following cosmological parameters consistent
with the measurement of the cosmic microwave background by
\citet{Planck_Collaboration:2016}:
the total matter density $\Omega_{m} = 0.3156$, 
the baryon density $\Omega_{b} = 0.0492$, 
the cosmological constant $\Omega_{\Lambda} = 0.6844$, 
the present-day Hubble parameter $H_0=67.27\, \mathrm{km/s/Mpc}$,
the spectral index of initial curvature perturbations $n_s=0.9645$,
and the amplitude of initial perturbations at $k=0.05\, \mathrm{Mpc}^{-1}$ 
being $A_{s} = 2.2065 \times 10^{-9}$.
For the $f(R)$ gravity run, we choose the value of $f_{R0}$ as
$|f_{R0}|=10^{-5}$ and set the functional form of $f(R) \propto
R/(R+\mathrm{const})$.  It is worth noting that the present-day linear
mass variance smoothed at $8\, h^{-1}\mathrm{Mpc}$, referred to as
$\sigma_8$, are set to 0.831 and 0.883 for the $\Lambda$CDM and $f(R)$
gravity runs, respectively.

From the simulation outputs, we identify dark matter halos with a
phase-space halo finder of \texttt{ROCKSTAR}
\citep{2013ApJ...762..109B}.  In this paper, we use the standard
definition for the halo mass, $M_{200}$, defined by a sphere with a
radius $r_{200}$ within which the enclosed average density is 200
times the mean mater density \citep{Navarro:1995iw}. We then create the 
halo catalogs with the mass threshold of $M_{200} > 10^{13}\, h^{-1}\,
M_{\odot}$ for the $f(R)$ gravity run, which contain 18,055 and 8,481 halos at $z=0$ and $1$, respectively.
Adopting the same mass threshold for the $\Lambda$CDM run leads to the lower number density, by $\lesssim 10\%$ \citep{Schmidt:2008tn,Li:2012}. Thus, to avoid inducing spurious distinguishability between the $\Lambda$CDM and $f(R)$ gravity models due to the different abundances, we include halos slightly lighter than $10^{13}\,h^{-1}M_{\odot}$ so that the number density of halos in the two runs
coincides.
The lowest halo mass can be resolved with 485 particles
in our simulation.  Our halo sample does not include subhalos.  The
velocity of each halo is computed by the average particle velocity
within the innermost 10\% of the virial radius.  
The principle axes of each halo in a projected plane are computed by diagonalizing the second moments of the distribution of all the member particles projected on the celestial plane, 
$I_{ij} \propto \sum_{k} \Delta x^i_k \Delta x^j_k$, where $\Delta x_k^i $ is the $i$th spatial component of the vector $\Delta \bfx_k $,
the difference between the positions of the halo center and $k$-th member particles, and the sum is over all the particles within the virial radius of the halo.
The two ellipticity components of each halo, projected along the third axis for instance, are estimated as \citep{Valdes:1983,Miralda-Escude:1991,Croft:2000gz}
\be
\gamma_{(+,\times)} = \frac{1}{I_{11}+ I_{22}} (I_{11}-I_{22} , 2I_{12}).
\ee

In measuring the redshift-space density field and the projected shape field, we rotate the simulation box and regard each direction along the three axes of the box as the line of sight. We thus have three realizations for each of the $\Lambda$CDM and $f(R)$ simulations, though they are not fully independent. Thus, in the following, all the quantities are averaged over the three projections.

%%% Figure1
\begin{figure}

\begin{center}
\includegraphics[width = 0.39\textwidth]{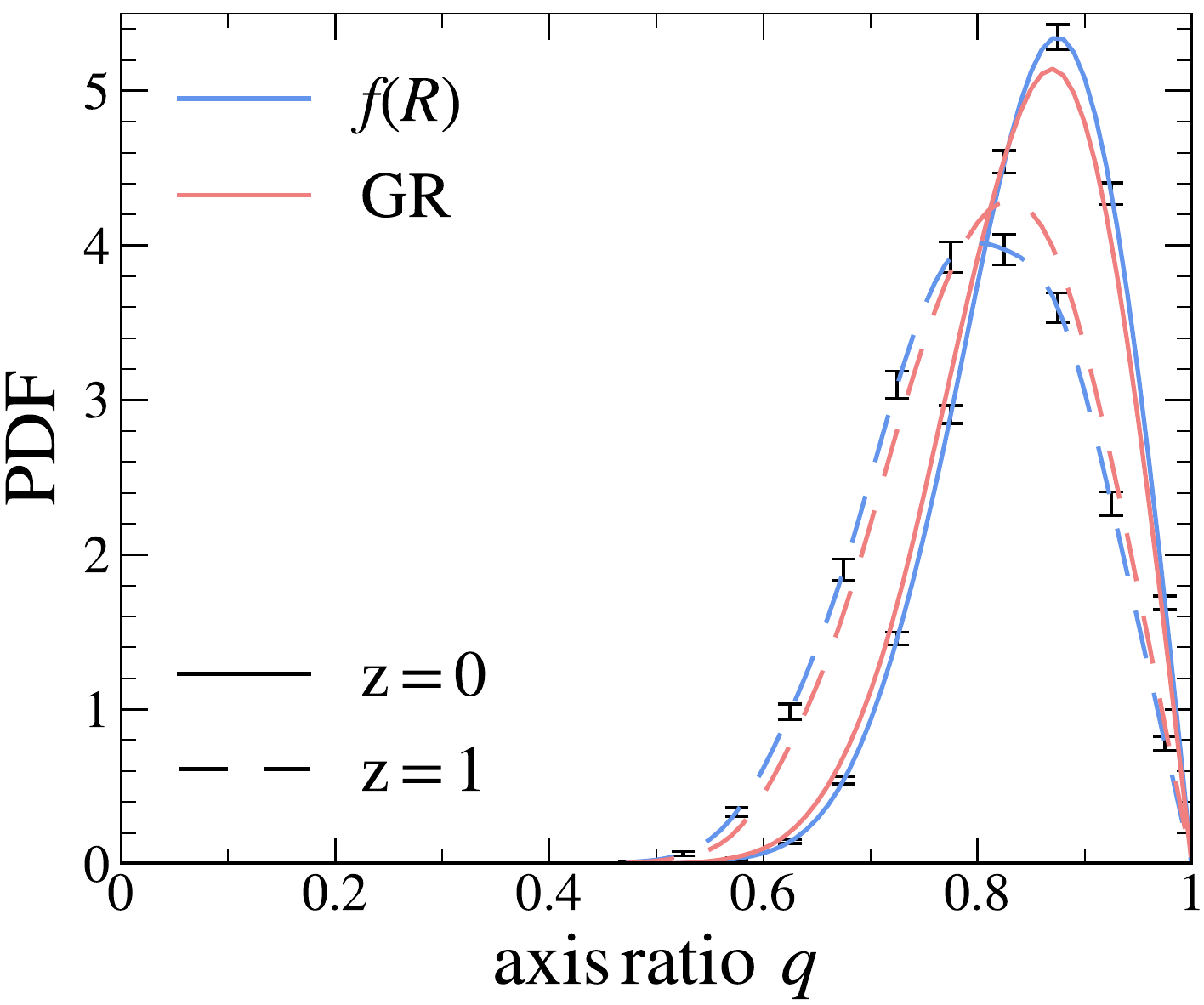}
\caption{
Distributions of minor-to-major axis ratios of projected halo shapes for $\Lambda$CDM and $f(R)$ gravity simulation runs. 
The solid and dashed curves show the results at $z=0$ and $1$, respectively. 
Halos with $M>10^{13}\,h^{-1}\,
M_{\odot}$ in the $f(R)$ gravity model are counted, and the lower mass threshold is adopted for $\Lambda$CDM halos to make the number densities equivalent with the former. 
The error bars shown only for the $f(R)$ gravity results represent the Poisson error. 
}
\label{fig:axis_factor}
\end{center}
\end{figure}

%%% Figure2
\begin{figure*}

\begin{center}

\includegraphics[width = 0.99\textwidth]{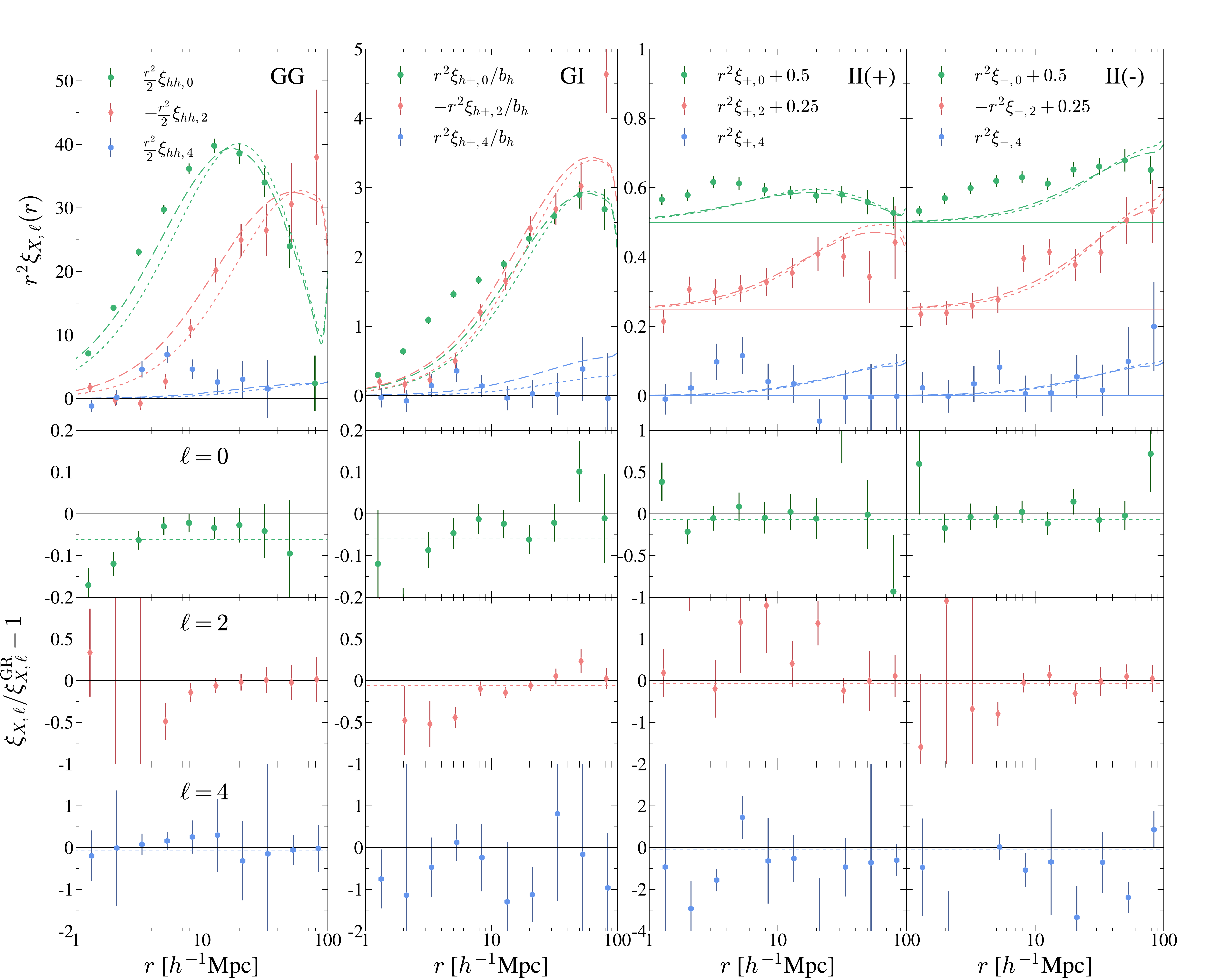}
\caption{ Multipole components of halo correlation functions in
  redshift space from $f(R)$ gravity mode at $z=0$, $\xi_{hh,\ell}$,
  $\xi_{h+,\ell}$, $\xi_{+,\ell}$, and $\xi_{-,\ell}$ from left to
  right panels in the first row.  The second row shows the monopole
  correlation function from the $f(R)$ gravity model shown in the
  first row divided by that from the $\Lambda$CDM model.  The third
  and fourth rows are similar with the second row but for the
  quadrupole and hexadecapole, respectively.  The dotted horizontal
  lines are the best-fitting value of
  $A=\xi_{X,\ell}/\xi_{X,\ell}^{\GR}$, where $A_{\GG} =0.938$, $A_{\GI} = 0.942$, and $A_{\II} = 0.930$. The dotted curves in the top
  row show $\xi_{X,\ell}=A\xi_{X,\ell}^{\GR}$, where
  $\xi_{X,\ell}^{\GR}$ is calculated from the LA model for
  demonstration. For another demonstration, 
  the dashed curves in the top row also show the LA model prediction 
  with the matter power spectrum $P_m^R$ being 
  calculated for the $f(R)$ gravity model using the \texttt{MGCAMB} code.}
\label{fig:2pcf}
\end{center}
\end{figure*}

Fig. \ref{fig:axis_factor} shows the distributions of minor-to-major axis ratios of projected halo shapes at $z=0$ and 1 for $\Lambda$CDM and $f(R)$ gravity simulation runs. As mentioned above, the distributions are averaged over the three realizations. 
By comparing the two distributions at $z=1$, one can see that halo shapes in $f(R)$ gravity are more elongated than those in $\Lambda$CDM. It could be because more masses undergo infalling along the filaments into halo centers in $f(R)$ gravity than in $\Lambda$CDM. 
Toward $z=0$ where the structure growth becomes more nonlinear, the difference of the axis ratio distributions between $f(R)$ gravity and $\Lambda$CDM models becomes less prominent. 
These trends at $z=0$ and $1$ are totally consistent with the earlier finding of \cite{LHuillier:2017pdi}.
In the following sections, we analyze the halo catalogs of $z=0$ only.

\section{Correlation functions in $f(R)$ gravity} \label{sec:cf}
\subsection{Correlation function measurements}\label{subsec:cf_results}

Here we measure the GG, GI and II statistics in the $N$-body
simulations.  Since all these statistics have explicit angular
dependences \citep{Hamilton:1992zz,Okumura:2020}, we consider their
multipole expansions in terms of the Legendre polynomials,
$\mathcal{P}_{\ell}$,
\begin{equation}
    \xi_{X,\ell}(r)= \frac{2\ell+1}{2} \int_{-1}^{1} \xi_X(\bfr)\mathcal{P}_{\ell}(\mu_{\bfr})d\mu_{\bfr},
\end{equation}
where $X=\{hh,h+,+,- \}$ and $\mu_{\bfr}=\hat{\bf r}\cdot \hat{\bfx}$
with a hat denoting a unit vector.  In linear perturbation theory all
the four correlation functions, $\xi_{X,\ell}$, have non-zero values
only for $\ell=0, 2$ and $4$.  We thus consider only these three
multipoles for each correlation function, and hence the number of the
total statistics is $4\times3=12$. 

Our estimators for the multipoles, $\xi_{X,\ell}$, are expressed as
\citep{Okumura:2020a}
\begin{equation}
    \xi_{X,\ell}(r) = \frac{2\ell+1}{2} \frac{1}{RR(r)} \sum\limits_{j,k|r=|\bfx_{k}-\bfx_{j}|} W_{X,jk}\mathcal{P}_{\ell}(\mu_{jk}) ,
\end{equation}
where $RR$ is the pair counts from the random distribution, which can
be analytically computed because we place the periodic boundary
condition on the simulation box.  For the GG, GI and II correlations,
$W_{hh,jk} = 1-\delta_{\ell 0}^K RR(r)/DD(r)$ , $W_{h+,jk} = \gamma_+(\bfx_j)$, and
$W_{\pm,jk}=\gamma_+(\bfx_j)\gamma_+(\bfx_k)\pm
\gamma_\times(\bfx_j)\gamma_\times(\bfx_k)$, respectively, where $\gamma_{(+,\times)}$ is redefined relative to the separation vector $\bfr$ projected on the celestial sphere, $\delta_{\ell\ell'}^K$ is the Kronecker delta and $DD(r)$ is the pair counts of halos at given separation $r$.
Clustering of halos in $f(R)$ gravity has been investigated in literature using $N$-body simulations \citep{Armalte-Mur:2017,Hernandez-Aguyo:2019,Alam:2021a,Garcia-Farieta:2021} and hydrodynamical simulations \citep{Arnold:2019}.
On the other hand, we will present the first measurements of intrinsic alignments from modified gravity simulations.

From the left to right of the first row in Fig. \ref{fig:2pcf}, we
show the multipole moments of the GG, GI, II($+$) and II($-$) correlation
functions measured from the $f(R)$ gravity simulations at $z=0$.  
Since we have three realizations for each simulation run by rotating the simulation 
box (see Section \ref{sec:f(R)gravity}), for each statistic we show the average over the three measurements. 
We adopt 10 logarithmically spaced bins in $r$ over $1<r<100[\mpcoh]$.
The data at the scales below and above this range are severely
affected by the halo exclusion effect and cosmic variance,
respectively.  The second, third and last rows of Fig. \ref{fig:2pcf}
are respectively the monopole, quadrupole and hexadecapole moments
from the $f(R)$ gravity simulation divided by those from the
$\Lambda$CDM simulation, $\xi_{X,\ell}/\xi^{\GR}_{X,\ell}$.
Note again that, all the correlation functions are measured in
redshift space which are a direct observable in real galaxy surveys,
and thus they are affected by RSD. 

Since the number density in the $f(R)$ gravity model
tends to be higher than that in the GR model, halos in $f(R)$ gravity
are less biased than those in $\Lambda$CDM with the same masses. 
To avoid this, we made our halo sample have the same number densities by lowering the mass threshold for the $\Lambda$CDM halos, as described in section \ref{sec:f(R)gravity}. Nevertheless,
the amplitude of the GG correlation in $f(R)$ gravity becomes
lower than that in GR as seen in the
lower panels of the leftmost column in Fig. \ref{fig:2pcf},
confirming the earlier finding that halos in $f(R)$ gravity tend to be less biased than those in $\Lambda$CDM \citep[e.g.,][]{Alam:2021a}.

Similarly to the GG correlation, the three multipoles of the GI
correlation in the $f(R)$ gravity model shows a negative deviation
from those in $\Lambda$CDM. Note that the non-zero hexadecapole comes
from the RSD effect \citep{Okumura:2020}.  Since the II($+$) and II($-$)
correlation functions are noisier than the GI correlation function, as
is known for the $\Lambda$CDM case, it is harder to see the difference
in the II correlation.  Nevertheless, for the monopole and quadrupole moments
one can see the same trend that the amplitude in $f(R)$ gravity is
smaller than that in $\Lambda$CDM, due to the fact that the amplitude
of IA, often characterized by $b_K$ (see equation \ref{eq:gamma_la}),
is positively correlated with the halo bias
\citep[e.g.,][]{Jing:2002,Okumura:2020a,Akitsu:2021,Kurita:2021}; more massive halos tend to be more
strongly aligned with the large-scale structure.

\subsection{Covariance matrix}

To perform a reliable statistical analysis, it is crucial to construct an accurate covariance error matrix. 
One of the most robust ways to do this is to use $N$-body simulations to generate many mock catalogs. It is, however, computationally very expensive as we need to run more than hundreds of independent simulations.
It is particularly impractical for the case of modified gravity simulations, since they are computationally even more expensive than standard $\Lambda$CDM simulations. 
We thus need to employ an alternative approach that is not so accurate as generating many mock catalogs based on $N$-body simulations. 

In our analysis, we use a bootstrap-resampling technique
\citep{Barrow:1984} to estimate a covariance error matrix
for the measured statistics in $f(R)$ gravity simulations.  With a
given original halo catalog, we construct a new catalog by randomly
choosing the same number of halos as the original catalog, allowing
repetition. We repeat this process until we obtain the required number
of bootstrap realizations, $N_{\bs}$.  For the $k$th realization
($1\leq k\leq N_{\bs}$), we measure the correlation function
multipoles, $\xi_{X,\ell}^{k}(r)$, where $X=\{hh,h+,+,-\}$ and
$\ell=0,2,4$.
Once again, $\xi_{X,\ell}^k$ is the average of the three measurements by rotating the simulation box for the $k$th realization.

\begin{figure}
\begin{center}
\includegraphics[width = 0.46\textwidth]{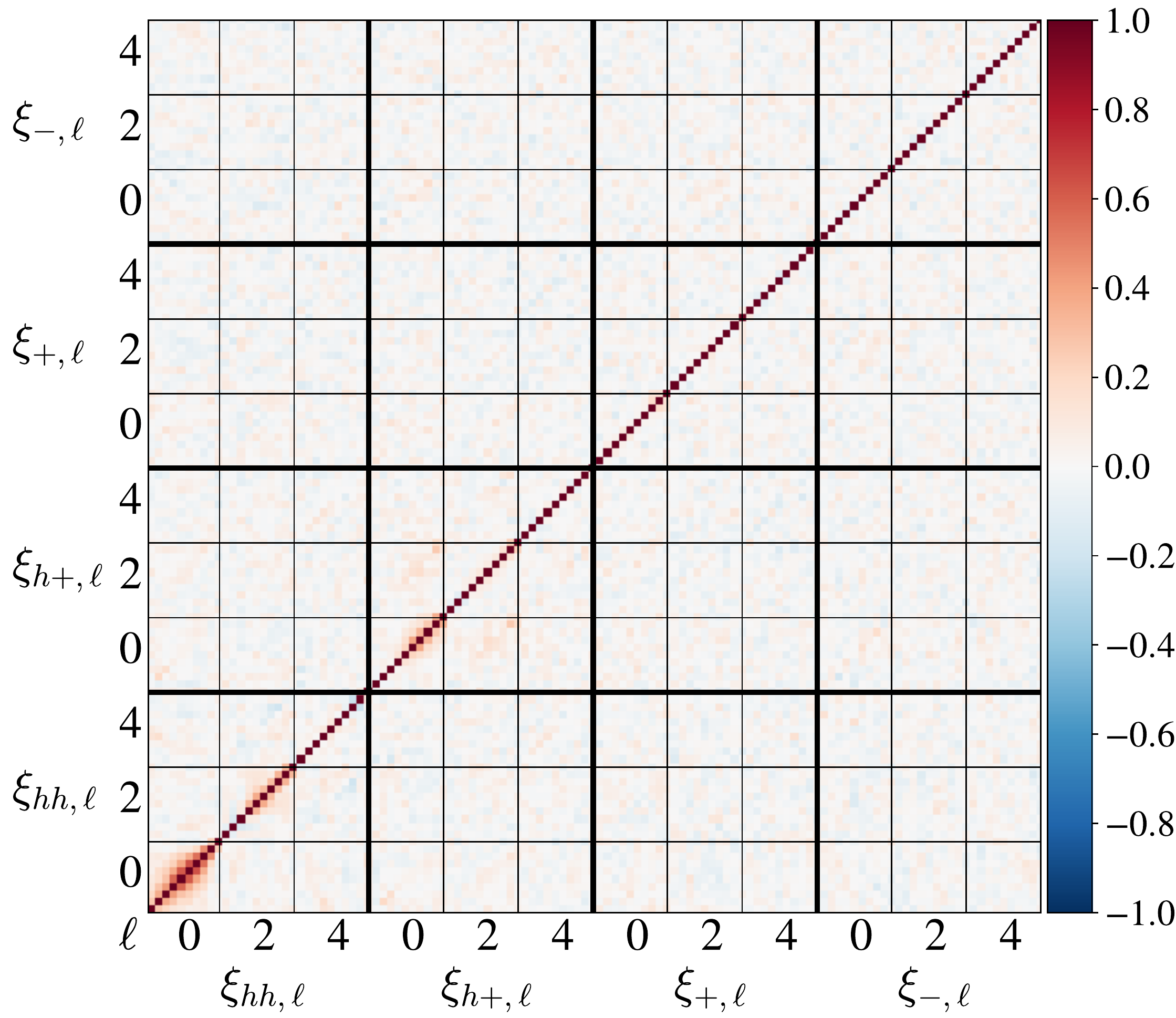}
\caption{ The $120\times 120$ normalized covariance matrix of the 12
  correlation function multipoles, each of which has 10 logarithmic
  separation, $\xi_{X,\ell}(r)$, measured at $z=0$. We use 500 bootstrap realizations, and for every realization we measure the statistics three times by rotating the simulation box.
  \label{fig:covariance}
}
\end{center}
\end{figure}

Given the measurements of the correlation functions, their covariance
matrix, ${\rm C}_{ij}^{X_\ell X'_{\ell'}}\equiv {\rm C}\left[
  \xi_{X,\ell}(r_i), \xi_{X',\ell'}(r_j) \right]$ can be estimated as
\begin{equation}
\begin{aligned}
    {\rm C}_{ij}^{X_\ell X'_{\ell'}} = \frac{1}{N_{\bs}-1}\sum\limits^{N_{\bs}}_{k}
    &\left[\xi^{k}_{X,\ell}(r_{i})-\bar{\xi}_{X,\ell}(r_{i})\right] \\
    &\times\left[\xi^{k}_{X',\ell'}(r_{j})-\bar{\xi}_{X',\ell'}(r_{j})\right],
    \label{eq:cov}
\end{aligned}
\end{equation}
where $\bar{\xi}_{X,\ell}$ is the average of $\xi_{X,\ell}^{k}$ over
$N_{\bs}$ realizations,
$\bar{\xi}_{X,\ell}=N_{\bs}^{-1}\sum^{N_{\bs}}_{k=1}\xi_{X,\ell}^{k}$.
Since the number of bins of each statistic is 10, the size of the full
covariance becomes $120\times 120$.  In this work, we choose
$N_{\bs}=500$ to avoid the covariance matrix being singular.  The
obtained full covariance matrix normalized by the diagonal elements,
${\rm C}_{ij}/({\rm C}_{ii}\cdot{\rm C}_{jj})^{1/2}$, is shown in
Fig. \ref{fig:covariance}.  The diagonal components, ${\rm
  C}_{ii}^{1/2}$, are shown as the errorbars of our statistics in
Fig. \ref{fig:2pcf}. 

%%% Figure3

Note that the bootstrap method is known to 
underestimate the cosmic variance \citep{Norberg:2009}.
Thus, the constraints we will obtain in the next section become
tighter than they should be. 
The purpose of this work is, however, to see how much adding the IA effect improves the distinguishability of different gravity models compared to the galaxy clustering analysis only, rather than the distinguishability itself. 
Thus, we do not expect that our conclusion is affected by the underestimation of the covariance matrix.

\section{Results and discussion}\label{sec:results}
\subsection{Distinguishing gravity models with IA }\label{subsec:distinguishability}

Here we investigate how well one can improve the distinguishability
between the $\Lambda$CDM and $f(R)$ gravity models by considering the
IA statistics.  For this purpose, we introduce a parameter $A$ which
characterizes the difference between the given statistics $\xi_X$ of
the two models, as
$A\equiv\xi_X({\bfr})/\xi_X^{\GR}({\bfr})=\xi_{X,\ell}(r)/\xi_{X,\ell}^{\GR}(r)$,
where $X=\{ hh,h+,+,-\}$ and $\ell=\{0,2,4\}$, and constrain the
parameter.  We add subscripts to $A$ depending on which statistics to
be used for the constraints, e.g., $A_{\GG}$, $A_{\GI}$ and $A_{\II}$
for the GG-, GI- and II-only analyses, respectively, and
$A_{\GG+\GI+\II}$ for their combination.  We adopt a simple $\chi^{2}$
statistic to constrain the parameter $A$ which is given by
\begin{align}
\chi^{2}(A) = \sum\limits_{i,j,\ell,\ell',X,X'} & 
\Delta^{X_\ell}_{i}
\left({\rm C}^{-1}\right)_{ij}^{X_\ell X'_{\ell'}} 
\Delta^{X'_{\ell'}}_{j}, \label{eq:chi2_multi}
\end{align}
where $\Delta^{X_\ell}_{i}= \xi_{X,\ell}(r_{i}) - A
\xi^{\GR}_{X,\ell}(r_{i})$. The covariance of the correlation
functions in $f(R)$ gravity, ${\rm C}_{ij}^{X_\ell X'_{\ell'}}$, is
the $120\times 120$ matrix, and for the single-statistics analysis of
the GG or GI correlation, the covariance is reduced to a $30\times 30$
submatrix, while the analysis of the II correlation, II($+$) and
II($-$), needs the $60\times 60$ submatrix.  Table \ref{tab:table}
summarizes the degree of freedom for each choice of the statistics.
Note that the constant model above is too simple and in reality the
deviation of $A$ from unity is scale-dependent. 
We perform a qualitative investigation of the scale dependence using 
a simple model in $f(R)$ gravity in the next subsection.
To properly take into account the scale dependence, however, the detailed 
modeling of IA statistics in modified gravity models is required. Furthermore, the amount of the
deviation from unity is not necessarily equivalent between different
statistics.  Thus, we do not focus on the best-fitting values of $A$
but are rather interested in how well we can exclude the possibility
of the correlation functions under the two models being equal, namely
$A=1$, and whether the constraint gets tighter by combining the IA
statistics with the clustering statistics.

\begin{figure}
\begin{center}
\includegraphics[width = 0.43\textwidth]{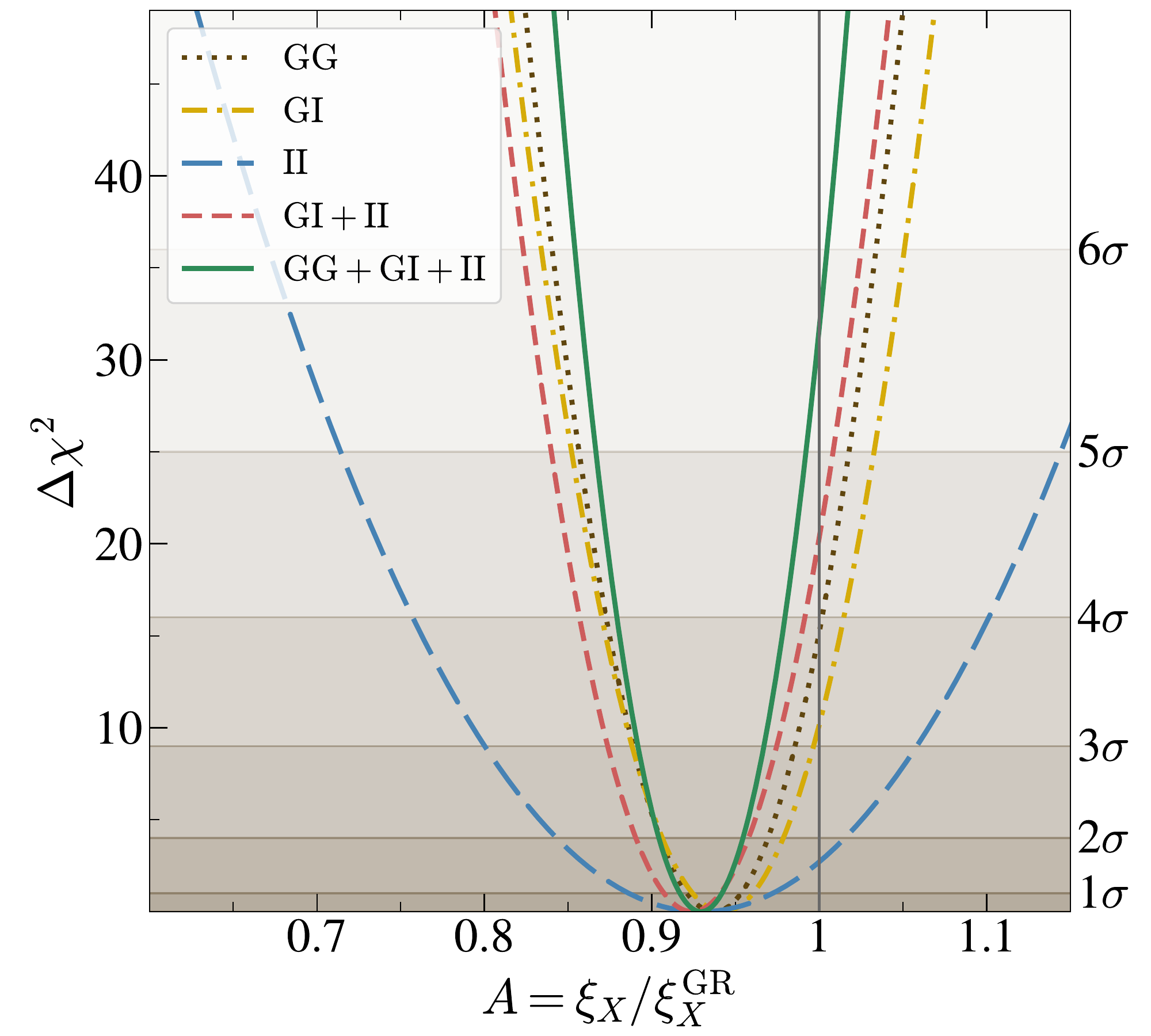}
\caption{ $\Delta\chi^{2}$ distribution for deviation parameter
  $A=\xi_{X}/\xi_X^{\GR}$ at $z=0$.  The horizontal lines indicate
  confidence levels as shown in the right y-axis.
  \label{fig:chisq}
}
\end{center}
\end{figure}

Fig. \ref{fig:chisq} shows $\Delta\chi^2=\chi^2-\chi^2_{\min}$ as a
function of $A$, where $\chi^2_{\min}$ is the minimized $\chi^2$ value
with the best-fitting $A$.  The black-dotted, yellow-dot-dashed and
blue-long-dashed curves are the constraints from the GG, GI and II
correlations, respectively. 
While the GG correlation gives the
tightest constraint as expected, the GI and II correlations also
provide meaningful constraints.
The best-fitting parameter $A_{\GG}$
is shown as the horizontal lines in the second, third and fourth rows
on the leftmost column of figure \ref{fig:2pcf}.  Similarly, $A_{\GI}$
is shown on the second leftmost column and $A_{\II}$ is on the third
and fourth columns.

%%%table1
\begin{table}
\begin{center}
\caption{Summary of constraints on the ratio of the statistics between
  $f(R)$ gravity and $\Lambda$CDM models, $A=\xi_X/\xi_{X}^{\GR}$, at
  $z=0$. The second column represents the degree of
  freedom for each constraint.}
\begin{tabular} {l c  c  c  cccccc} 
\hline
 
 Statistics & d.o.f. $\nu$ & $\chi_{\min}^2$ & $\Delta\chi^2(A=1)$  (C.L.)  \\ [0.1ex] 
 \hline
 GG & 29 & 28.15 & $15.2$  $(3.9\sigma)$ \\ 
 GI & 29 & 42.9 & $10.17$  $(3.19\sigma)$ \\
 II & 59 & 41.87 & $2.71$  $(1.6\sigma)$ \\
 GI$+$II & 89 & 83.98 & $20.35$  $(4.51\sigma)$ \\ 
 GG$+$GI$+$II & 119 & 114.64 & $31.74$  $(5.63\sigma)$ \\[.1ex] 

 \hline
\label{tab:table}
\end{tabular}
\end{center}
\end{table}

%%% Figure4

%%% Figure5
\begin{figure*}
\begin{center}
\includegraphics[width = 0.85\textwidth]{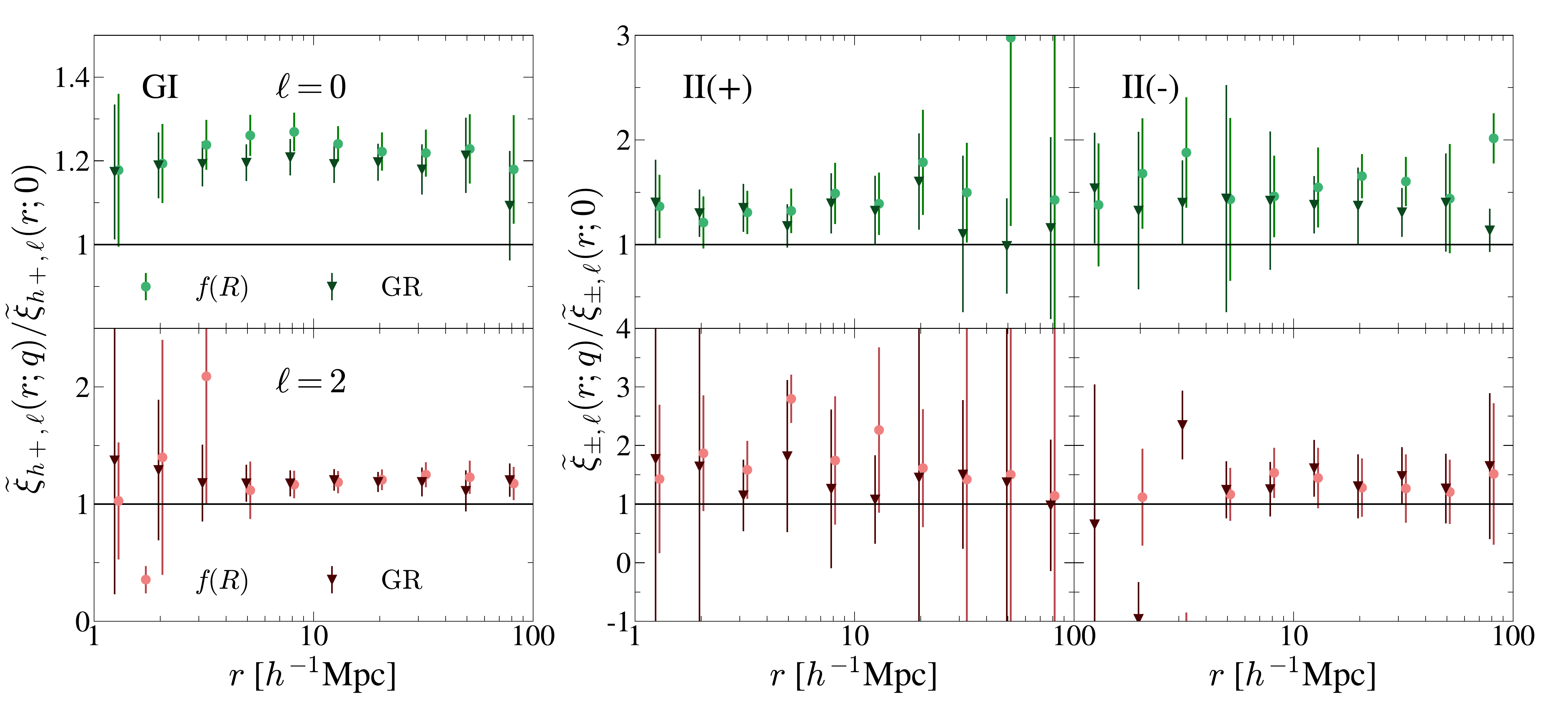}
\caption{ Left-hand set: ratios of normalized GI correlation functions with $q>0$ and $q=0$, $\wt{\xi}_{h+}(r;q)/\wt{\xi}_{h+}(r;0)$ (see Eq. (\ref{eq:normalized_gi}) for the definition). Right-hand set: Same as the left-hand set but the ratios of normalized II correlation functions, $\wt{\xi}_{\pm}(r;q)/\wt{\xi}_{\pm}(r;0)$ (Eq. (\ref{eq:normalized_ii})). The upper and lower rows show results for the monopole and quadrupole moments, respectively. The deviation of the ratio from unity indicates the correlation between the halo ellipticity and orientation. The results for the $f(R)$ gravity and $\Lambda$CDM models are respectively shifted in the horizontally positive and negative directions for clarity.
\label{fig:axis_ratio}}
\end{center}
\end{figure*}

We study how well one can improve the distinguishability
between the $\Lambda$CDM and $f(R)$ gravity models by combining the IA
statistics with the conventional clustering statistics.  The
constraint using the combination of GI and II correlations is shown as
the red curve in Fig. \ref{fig:chisq}.  It is interesting to see that
though the constraint from the conventional GG-correlation analysis is
stronger and excludes the case of $\xi_{hh,\ell}=\xi_{hh,\ell}^{\GR}$
with $3.9\sigma$ C.L., one can achieve the meaningful constraint from
GI+II ($4.51\sigma$).  Once we combine all the statistics, namely 12
multipoles, the distinguishability reaches $5.63\sigma$. Note that the
analysis is based on various assumptions and simplifications, and thus
these numbers do not have much importance. Rather, the amount of the
improvement ($\sim 40\%$) matters.  
All the obtained numerical values are summarized in table \ref{tab:table}.

\subsection{Linear alignment model in $f(R)$ gravity}

As a demonstration, we calculate the model prediction of the GG, GI
and II correlations in $f(R)$ gravity using linear perturbation
theory.  In Fourier space, the halo density field in redshift space,
$\delta_h(\bfk)$, is linearly related to the underlying matter density
field in real space, $\delta_m^R(\bfk)$, via the Kaiser factor
\citep{Kaiser:1987}, $\delta_h(\bfk)=(b_h+f\mu_{\bfk}^2)
\delta_m^R(\bfk)$, where $\mu_{\bfk}=\hat{\bfk}\cdot\hat{\bfx}$ and
$b_h$ is the linear halo bias.
The superscript $R$ denotes a real-space quantity. 
The the growth rate, $f$, is defined as $f(z)=-d\ln{D(z)}/d\ln{(1+z)}$, with $D(z)$ being the growth factor. 
Different gravity models predict different values of $D(z)$ and $f(z)$.
For the IA statistics, we adopt the linear
alignment (LA) model \citep{Catelan:2001,Hirata:2004}, which relates
the ellipticity field linearly to the tidal gravitational field,
described in Fourier space as \be \gamma_+({\bfk})=b_K(k_x^2 -
k_y^2)\delta_m^R({\bfk})/k^2, \label{eq:gamma_la} \ee where $b_K$ is
the shape bias parameter. Following \cite{Okumura:2020} we introduce
the expression,
\begin{equation}
    \Xi_{\ell}(r)= \frac{1}{2\pi^{2}}\int_{0}^{\infty}k^{2-n}dkP_m^R(k)j_{\ell}(kr),
\label{eq:Xi}
\end{equation}
where $P_m^R(k)$ is the matter power spectrum in real space. We can then write all
the IA statistics in the LA model in terms of $\Xi_{\ell}$, similarly
to the case of the GG correlation \citep{Hamilton:1992zz,Okumura:2020}. 
In the following, we compute $P_m^R(k)$ in two ways.

First, we use the \texttt{CAMB} code \citep{Lewis:2000} to compute $P_m^R(k)$ in the $\Lambda$CDM model and obtain the correlation functions, $\xi_{X,\ell}^{\GR}$. We then multiply it by the best-fitting value of $A$ obtained in Section \ref{subsec:distinguishability} to have the prediction for $f(R)$ halos, $\xi_{X,\ell}=A\xi_{X,\ell}^{\GR}$. 
The bias parameter $b_h$ for the halos in $\Lambda$CDM is determined
as $b_h^{\mathrm{GR}}=1.40$ by fitting the ratio of the GG correlation function to
the matter correlation function $\xi_m^R = \Xi_0$,
$b_h^2=\xi_{hh,0}^R / \xi_m^R$, to the measurement in the $\Lambda$CDM simulation on large scales.
Similarly, $b_K$ is determined as $b_K^{\mathrm{GR}}=0.46$ by fitting $b_K^2=(15/8)\xi_{+,0}^R / \xi_m^R$.

Second, to take into account the scale dependence induced by the modification of gravity, we compute $P_m^R(k)$ directly in the $f(R)$ gravity model using the \texttt{MGCAMB} code \citep{Zhao:2009,Hojjati:2011,Zucca:2019xhg}. Similarly to the first case above, we determine the two biases, $b_h$ and $b_K$, by taking the ratios of the real-space correlation functions of halos measured from the $f(R)$ gravity simulation and the matter correlation function from the \texttt{MGCAMB}. 
They are determined as $b_h=1.34$ and $b_K=0.42$, lower than the values for the $\Lambda$CDM model, as expected. 
Different gravity models predict different values of the growth rate, $f(z)$. Furthermore, $f(z)$ can become scale dependent in modified gravity models, 
which arise from the effective gravitational constant \citep{Narikawa:2010}. Thus, the GG and GI correlation functions induce a further scale dependence due to $f(z)$ in the Kaiser factor.

In the top row of Fig. \ref{fig:2pcf}, we show the two LA predictions for halo statistics in $f(R)$ gravity explained above, one with the correlation function in $\Lambda$CDM multiplied by the best-fitting parameter $A$ obtained in section \ref{subsec:distinguishability} (dotted curves), and another with the correlation function in $f(R)$ gravity directly computed using the \texttt{MGCAMB} code (dashed curves). Note that they are not fitting results of the LA model predictions: the parameter $A$ is constrained by the measured ratio of $\xi_{X,\ell}/\xi_{X,\ell}^{\GR}$ and the bias parameters are also determined by the simulation measurements. 
Nevertheless, we can see the LA model qualitatively explains the measured correlation functions of halos in $f(R)$ gravity. 
A close look at the small scale behavior of the correlation functions shows that the prediction based on the \texttt{MGCAMB} gives better agreement, particularly for the GG and GI correlation functions. 
It is important to note that the difference between the two model curves comes from the scale dependence of the growth factor $D(z)$ and its growth rate $f(z)$ which was not considered in section
\ref{subsec:distinguishability}. 
Properly taking into account the effect will help improve the distinguishability between $\Lambda$CDM and $f(R)$ gravity models.

\subsection{Correlation of halo shape and its orientation}
Is there additional information encoded in the halo shape and orientation to distinguish different gravity models? To answer this, we consider normalized alignment statistics, which were introduced in \cite{Okumura:2009a}. The normalized GI and II correlation functions are defined as
\begin{align}
\wt{\xi}_{h+}(\bfr;q) &= \left\langle \frac{1-q^2}{1+q^2} \right\rangle^{-1}\xi_{h+}(\bfr;q), \label{eq:normalized_gi} \\
\wt{\xi}_{\pm}(\bfr;q) &= \left\langle \frac{1-q^2}{1+q^2} \right\rangle^{-2}\xi_{\pm}(\bfr;q), \label{eq:normalized_ii} 
\end{align}
where $\left\langle \frac{1-q^2}{1+q^2} \right\rangle$ is the value averaged over all the halos used for our analysis and $\xi_X(\bfr;q)$ 
are the same as $\xi_X(\bfr)$ in equation (\ref{eq:statistics}) but the dependence on $q$ is explicitly written. 
These normalized alignment statistics are useful because if there is no correlation between axis ratios and orientations, we simply expect $\wt{\xi}_{X}(\bfr;q) = \wt{\xi}_{X}(\bfr;0)$. Namely, even though the axis-ratio distributions were different between two models as shown in figure \ref{fig:axis_factor}, it would not affect the distinguishability of gravity models if the normalized statistics were used. 

In Fig. \ref{fig:axis_ratio}, we show the ratio of the multipoles, $\wt{\xi}_{X,\ell}(r;q) / \wt{\xi}_{X,\ell}(r;0)$. In all the statistics the ratios tend to be greater than unity, which implies that there are non-negligible correlations between the halo shape and orientation, consistent with the finding of \cite{Okumura:2009a}. Interestingly, for most of the statistics, the correlation is stronger in $f(R)$ gravity than in $\Lambda$CDM. 
This indicates that if properly modeled, considering the correlation between halo shapes and orientations potentially helps improve the distinguishability between $\Lambda$CDM and $f(R)$ gravity models. \cite{Blazek:2011} studied the correlation in $\Lambda$CDM based on the LA model. We leave the study of the correlation between halo ellipticity and orientation for the $f(R)$ gravity model as future work.

\section{Conclusions} \label{sec:conclusion}
In this paper, we have presented the first measurements of IA in a
gravity model beyond GR using the two types of IA statistics, the GI
and II correlation functions of halo shapes from $f(R)$ gravity
simulations. By comparing them with the same statistics measured in
$\Lambda$CDM simulations, we found that the IA statistics in different
gravity models show distinguishable features, with a trend similar to
the case of conventional galaxy clustering statistics.
Quantitatively, IA statistics enhance detectability of the imprint of
$f(R)$ gravity on large scale structures by $\sim 40\%$ when combined
with the conventional halo clustering.
We also found that the correlation between the axis ratio and orientation 
of halos becomes stronger in $f(R)$ gravity than that in $\Lambda$CDM. 

Our constraints on different gravity models have been made assuming
that the effect of the modified gravity on the clustering and IA
statistics can be perfectly modeled. However, in the analysis of
actual observations of IA, one needs to model the present statistics
from linear to quasi non-linear scales. While there are several
theoretical studies of IA beyond linear theory in GR
\citep[e.g.,][]{Blazek:2019,Vlah:2020}, such predictions need to be
carefully tested and extended to gravity models beyond
GR. Furthermore, in real surveys, one observes shapes of galaxies, not
of halos, thus misalignment between the major axes of galaxies and
their host halos \citep{Okumura:2009} would degrade the detection
significance of IA even in a modified gravity scenario.  As a result,
the distinguishability between different gravity models would be
degraded compared to the results obtained in this paper.
On the other hand, in this work we used only the amplitude of the
multipole moments of the clustering and IA statistics, not the full
shape of the underlying matter power spectrum which contains ample
cosmological information but is more severely affected by the
nonlinearities.  Therefore, constraining power could eventually be
either enhanced or suppressed.  The more detailed and realistic
analysis of clustering and IA statistics beyond a consistency test of
GR will be performed in our future work.

\section*{Acknowledgments}
We thank Atsushi Taruya for sharing the code to calculate the growth rate under the $f(R)$ gravity model.
We also thank the anonymous referee for their careful reading and suggestions.
TO acknowledges support from the Ministry of Science and Technology of
Taiwan under Grants Nos. MOST 110-2112-M-001-045- and 111-2112-M-001-061- and the Career
Development Award, Academia Sinica (AS-CDA-108-M02) for the period of
2019-2023.  This work is in part supported by MEXT KAKENHI Grant
Number (18H04358, 19K14767, 20H05861). Numerical computations were in
part carried out on Cray XC50 at Center for Computational
Astrophysics, National Astronomical Observatory of Japan.

\section*{Data Availability}
The data underlying this article will be shared on reasonable request
to the corresponding authors.

%\bibliographystyle{mnras}
%\bibliography{refs}
\bibliography{ms.bbl}

\bsp	% typesetting comment
\label{lastpage}
\end{document}